\documentclass[twocolumn,superscriptaddress,letter]{revtex4}%
\usepackage{amssymb}
\usepackage{color}
\usepackage{graphicx}
\usepackage{dcolumn}
\usepackage{bm}
\usepackage[header,title,page,titletoc]{appendix}
\usepackage{amsmath}
\usepackage{amsfonts}%
\providecommand{\U}[1]{\protect \rule{.1in}{.1in}}
\begin{document}
\title{It from Knot}
\author{Su-Peng Kou}
\thanks{Corresponding author}
\email{spkou@bnu.edu.cn}
\affiliation{Department of Physics, Beijing Normal University, Beijing, 100875, P. R. China}

\begin{abstract}
Knot physics is the theory of the universe that not only unified all the
fundamental interactions but also explores the underlying physics of quantum
mechanics. In knot physics, the most important physical result is the
unification of everything (including matter, motion, interaction and
space-time, ...) into the entanglement between vortex-membranes: our universe
is a projection of a four three dimensional entangled vortex-membranes in five
dimensional inviscid incompressible fluid, of which the low energy effective
theory not only reproduces the Standard model -- an $\mathrm{SU}%
_{\mathrm{strong}}\mathrm{(3)}\otimes \mathrm{SU}_{\mathrm{weak}}%
\mathrm{(2)}\otimes$\textrm{U}$_{\mathrm{Y}}$\textrm{(1)} gauge theory but
also leads to the physics of general relativity. In a word, it from knot.

\end{abstract}
\maketitle

\section{Introduction}

In modern physics, quantum physics including quantum mechanics and quantum
field theory describes the universe. In the framework of quantum field theory,
it is believed that the interaction comes from exchanging virtual bosons via
local gauge symmetry principles. The Standard model (an
$\mathrm{SU_{\mathrm{strong}}(3)}\otimes \mathrm{SU}_{\mathrm{weak}%
}\mathrm{(2)}\otimes$\textrm{U}$_{\mathrm{Y}}$\textrm{(1)} gauge theory with
Higgs mechanism due to spontaneous symmetry breaking) is a special type of
quantum field theory that focuses on three non-gravitational interactions:
weak interaction, strong interaction, and electromagnetic
interaction\cite{stand}. On the other hand, to understand gravitational
interaction, the general relativity is developed by Einstein that provides a
unified description of gravity as a geometric property of space-time.

A \emph{Theory of Everything} (TOE) to unify all interactions is a dream of
physicists. There are several approaches towards TOE. String theory is a
possible theory of TOE\cite{ss}. According to string theory, quantum matter
consists of vibrating strings (or strands) and different oscillatory patterns
of strings become different particles with different masses. For the
researcheres on string theory, the key point can be summarized into "it from
string". In condensed matter physics, the idea of our universe as an
\textquotedblleft emergent\textquotedblright \ phenomenon has become
increasingly popular. In emergence approach, a deeper and unified
understanding of the universe is developed based on a complicated many-body
system with long range entanglement. This is the idea of "it from qubit".
Different quantum fields correspond to different many-body systems: the vacuum
corresponds to the ground state and the elementary particles correspond to the
excitations of the systems\cite{wen}. In addition, there exist many other
proposals of TOE from different points of view, such as loop quantum gravity
theory\cite{loop}, G. Lisi's \textrm{E8} theory\cite{e8}, C. Schiller's Strand
Model\cite{Strand}, R. Penrose's twistor theory\cite{twistor}, Grigory E.
Volovik's emergent theory in liquid helium\cite{vo}...

In this paper, we point out a new approach -- \emph{knot physics} towards
understanding our universe\cite{kou1,kou2,kou3}. This paper summarizes the
earlier three long papers, \cite{kou1,kou2,kou3}. In knot physics, our
universe is a projection of four entangled three-dimensional vortex-membranes
in five dimensional inviscid incompressible fluid. The collective motions of
this system are described by fermionic elementary particles (different types
of knots) and gauge fields. In particular, all four types of fundamental
interactions (electromagnetic, weak, strong and gravitational interactions)
are unified into single simple framework -- knot. That means, "it from knot".

\begin{figure}[ptb]
\includegraphics[clip,width=0.5\textwidth]{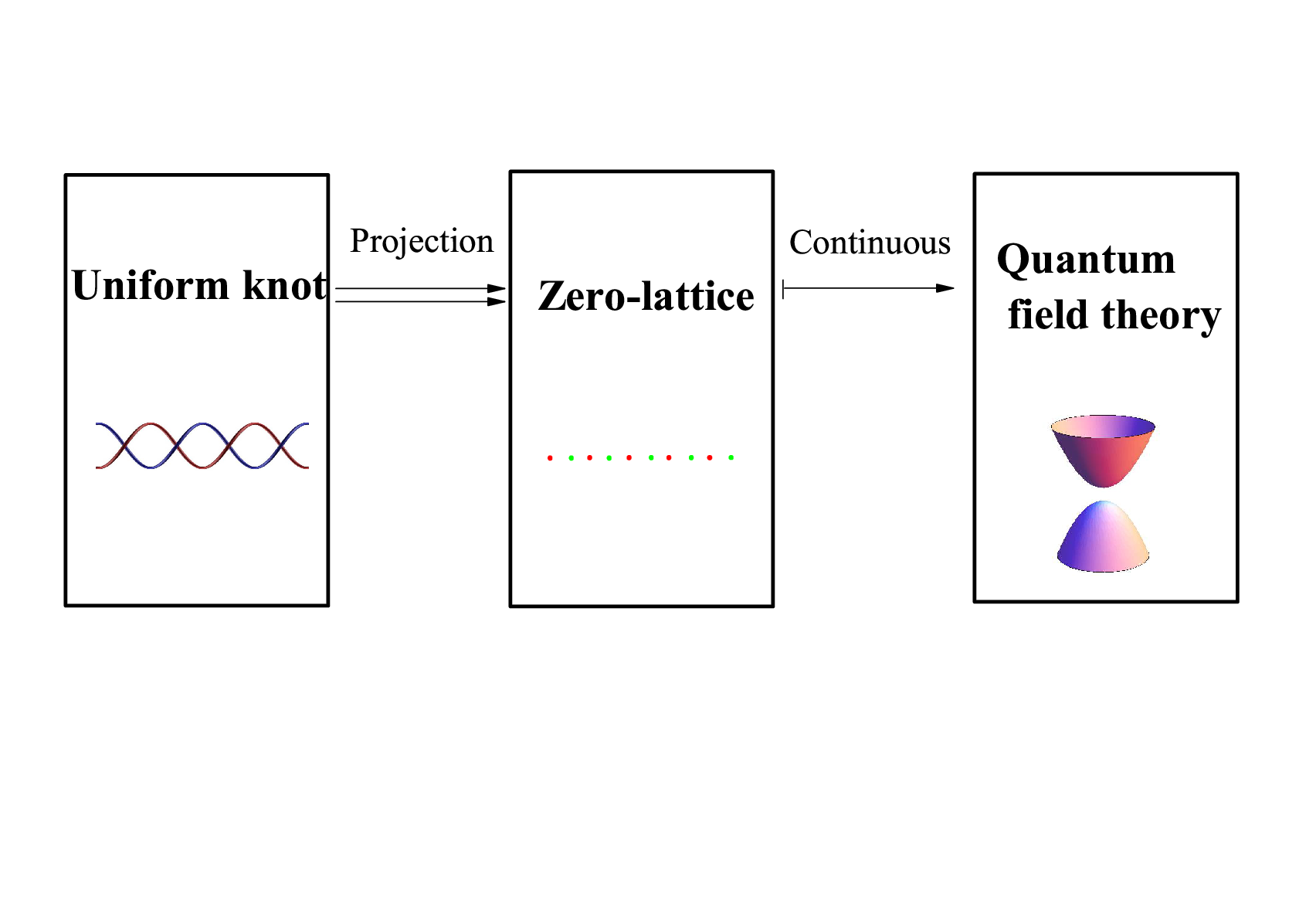}\caption{The framework
structure of knot physics}%
\end{figure}

\section{The foundation of knot physics}

Fig.1 is the framework structure of knot physics. In this figure, we take a
one dimensional (1D) knot with two entangled vortex-membranes as an example.
Our starting point is a knot Under projection, the knot turns into
zero-lattice. In the continuum limit, the effective theories of zero-lattice
become quantum field theories. We show the correspondence between modern
physics and knot physics as,
\begin{align*}
\text{The universe} &  \iff \text{ Entangled vortex-membranes }\\
&  \text{(A uniform knot),}\\
\text{The space-time} &  \iff \text{ 3+1D zero-lattice,}\\
\text{Matter} &  \iff \text{ Extra zeroes,}\\
\text{Motion} &  \iff \text{ The changing of the zero distribution,}\\
\text{Quantum mechanics} &  \iff \text{ Geometric Biot-Savart mechanics}\\
&  \text{for extra zeroes.}%
\end{align*}
Here, $\Longleftrightarrow$ denotes the correspondence between modern physics
and knot physics.

The following points are foundation of knot physics.

Vortices are extended objects with singular vorticity in divergence-free
inviscid incompressible fluid that can be regarded as a closed oriented
embedded subvortex-membrane with Marsden-Weinstein (MW) symplectic
structure\cite{leap}. For two dimensional (2D) inviscid incompressible fluid,
we have 0-dimensional point-vortices; For three dimensional (3D) case, we have
1D vortex-lines; For four dimensional case, we have 2D vortex-surfaces; For
five dimensional (5D) case, we have 3D vortex-membranes. The geometric
Biot-Savart equation for a 3D vortex-piece under local induction approximation
(LIA) can be described by Hamiltonian formula\cite{leap} and the Hamiltonian
of the vortex-membranes is just $3$-volume.

For Kelvin waves on a 3D helical vortex-membrane, the plane-wave is described
by a complex field, $\mathrm{z}(\vec{x},t)=r_{0}e^{\pm i\vec{k}\cdot \vec
{x}-i\omega t}$ where $\vec{k}$ is the winding wave vector along a direction
on 3D vortex-membrane with $\left \vert \vec{k}\right \vert =\frac{\pi}{a}$ and
$a$ is a fixed length that denotes the half pitch of the windings. For two
entangled vortex-membranes, the nonlocal interaction leads to leapfrogging
motion\cite{1}. For leapfrogging motion, the entangled vortex-membranes
exchange energy in a periodic fashion. The winding radii of two
vortex-membranes oscillate with a fixed leapfrogging angular frequency
$\omega^{\ast}$.

In the paper \cite{kou1,kou2,kou3}, a periodic entanglement-pattern between
vortex-membranes is called \emph{(uniform)} \emph{knot} that is described by a
special pure state of Kelvin waves with fixed wave length. A \emph{zero} on a
knot is a sharp, fragmentized, topological phase-changing in knot physics and
would become a "point" after projection. The definition of a "knot" is based
on periodic structures of zeroes that is similar to atom-crystal where the
atoms form a periodic arrangement: A zero corresponds to an elementary object
of a knot; A knot can be regarded as composite system of multi-zero.

\begin{figure}[ptb]
\includegraphics[clip,width=0.52\textwidth]{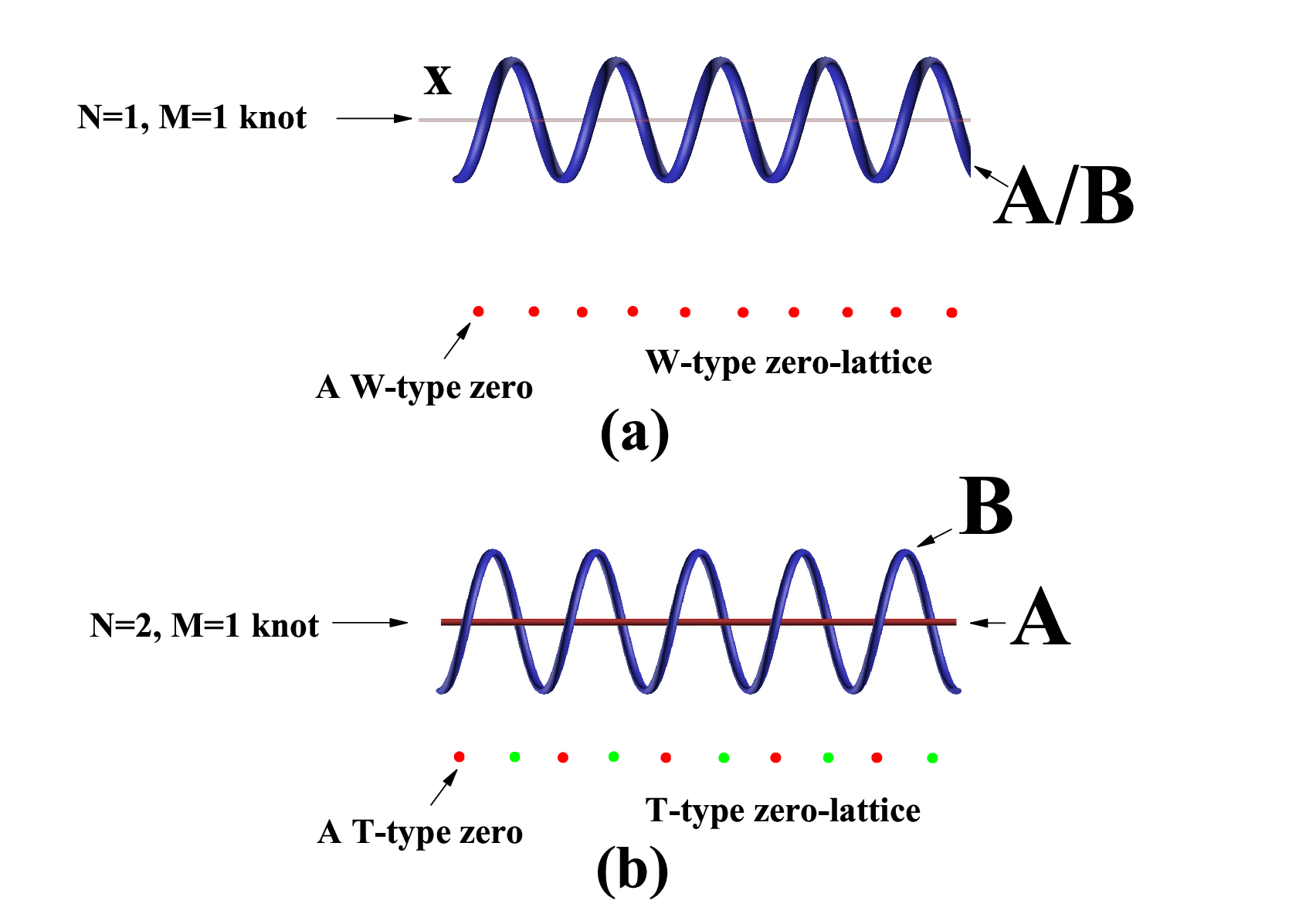}\caption{(a) 1D W-type
knot (two vortex-lines with winding central axis) and its projected
zero-lattice for winding cental axis; (b) 1D T-type knot (two entangled
vortex-lines with flat central axis) and its projected zero-lattice between
the vortex-lines}%
\end{figure}

To characterize the deformed knot, we \emph{project} the vortex-membranes
$\mathrm{z}(\vec{x},t)=\xi(\vec{x},t)+i\eta(\vec{x},t)$ by solving the zero
equation $\hat{P}[\mathrm{z}(\vec{x},t)]\equiv \xi(\vec{x},t)=0.$ After
projecting, we show that different entangled vortex-membranes correspond to
different\emph{ zero-lattices}. In knot physics, the information is
characterized by the distribution of zeroes. Here, the information unit
corresponds to a\emph{ zero} between projected vortex-membranes. There exist
two types of zero-lattices -- T-type zero-lattice and W-type zero-lattice. See
the illustration in Fig.2. Firstly, we consider the W-type zero-lattice from
projection on a central axis of the two vortex-membranes. A crossing between a
helical central axis $\mathrm{z}(\vec{x},t)$ and a straight mathematical
membrane ($\mathrm{z}(\vec{x},t)=0$) in its center corresponds to a solution
of the equation $\hat{P}_{\theta}[\mathrm{z}(\vec{x},t)]=0,$ that is $\xi
(\vec{x},t)=0.$ Next, we consider the T-type zero-lattice from projection on
two entangled vortex-membranes themselves. A crossing between the two
vortex-membranes ($\mathrm{z}_{\mathrm{A}}(\vec{x},t)$ and $\mathrm{z}%
_{\mathrm{B}}(\vec{x},t)$) corresponds to a solution of the equation $\hat
{P}[\mathrm{z}_{\mathrm{A}}(\vec{x},t)]=\hat{P}[\mathrm{z}_{\mathrm{B}}%
(\vec{x},t)],$ that is $\xi_{\mathrm{A}}(\vec{x},t)=\xi_{\mathrm{B}}(\vec
{x},t).$ According to the existence of two types of zero-lattices, there are
two types of zeroes: W-type zero and T-type zero (See Fig.2).

\section{Emergent quantum mechanics}

The elementary excitations of a knot are zeroes with \emph{unit information}.
For a (T-type or W-type) knot, there are three conserved physical quantities:
the energy of a (static) zero that is proportional to its volume on
vortex-membrane; the (Lamb impulse) angular momentum that is proportional to
its volume in the 5D fluid; the winding number $\pm \frac{1}{2}$ or the zero
number $\pm1$. According to the geometric character of the three conserved
physical quantities, the shape of zero will never be changed. However, the
"zero" can split and the three physical quantities are conserved for all
knot-pieces. Quantum mechanics becomes the mechanics to determine the
distribution of knot-pieces.

The five fundamental assumptions of emergent quantum mechanics in knot physics
are given by:

\begin{enumerate}
\item \textit{Quantum states}: To characterize the distribution of zeroes, we
have introduced the concept of "knot-pieces". Quantum mechanics is a mechanics
to determine the distribution of knot-pieces, that is the information of a
(deformed) knot. Therefore, an important issue is to obtain the spatial and
temporal distribution of the knot-pieces that has the information of a
(deformed) knot with an extra zero. Quantum state for a particle is just the
distribution of zero (or the distribution of knot-pieces for a deformed knot
with an extra zero). We point out that the distribution of the knot-pieces and
plays the role of the wave-function in quantum mechanics: the angle becomes
the quantum phase angle of wave-function and the zero density becomes the
density of knot-pieces.

\item \textit{Operators}: From the point view of information, the momentum
denotes the spatial distribution of knot-pieces (or information) and the
energy denotes the temporal distribution of knot-pieces (or information). From
the point view of fluid mechanics, the momentum is proportional to Lamb
impulse and the energy is proportional to the rotating velocity in extra
dimensions. As a result, the energy and momentum for the extra zero are
described by operators. The effective Planck constant is obtained as projected
angular momentum in extra dimensions of the extra knot.\ 

\item \textit{Schr\"{o}dinger equation}: The geometric Biot-Savart equation
for the deformed knot becomes Schr\"{o}dinger equation for probability waves
of the knot-pieces of a zero. And the classical functions for perturbative
Kelvin waves become wave-functions for quantum particle.

\item \textit{Fermionic statistics}: In quantum mechanics, another assumption
is statistics of Fermionic particles. In knot physics, when we exchange two
zeroes, the angle in extra dimensions (that is just the phase of
wave-function) is $\pi$. As a result, an zero becomes a fermionic particle.

\item \textit{Measurement theory}: The information is the distribution of
zeroes between projected vortex-membranes that is determined by zero density
for a given quantum state described by the wave-function. That means to
identify the deformation of vortex-membranes, people detect the distribution
function of knot-pieces. During measurement processes, the phase angle of
knot-pieces is fixed that depends on the time of measurement. This gives an
explanation of "wave-function collapse" in quantum mechanics.\emph{ }The
probability of emergent quantum mechanics comes from \emph{dynamically}
detecting the distribution of zeroes, or dynamically projecting entangled
vortex-membranes via stochastic projected time under "fast clock effect" that
leads to stochastic projected angle.
\end{enumerate}

In addition, we discuss the complementarity principle and wave--particle
duality in knot physics.

In knot physics, complementarity principle comes from complementarity property
of knots. On the one hand, a knot-piece is phase-changing -- a sharp,
time-independent, topological phase-changing; On the other hand, a knot-piece
has a fixed phase angle that is determined by wave function. One cannot
exactly determine the phase angle of a knot-piece by observing its phase-changing.

Wave-particle duality of quantum particles comes from fragmentation of a unit
information in knot physics. On the one hand, a knot is information unit that
is a sharp, fragmentized, topological phase-changing and would become a
"point" after projection. Thus, it shows particle-like behavior; On the other
hand, the distribution of knot-pieces shows wave-like behavior that is
characterized by wave-functions.

\section{Emergent gauge theory}

\begin{widetext}
\begin{table*}[t]%
\begin{tabular}
[c]{|c|cccc|}\hline Composite knot& Quantum field theory & & &\\
\hline   $\mathcal{N}=1$ and $\mathcal{M}=1$& Weyl Fermion model && &
\\ \hline $\mathcal{N}=2$ and $\mathcal{M}=1$& Dirac Fermion model & & &
\\ \hline $\mathcal{N}=4$ and $\mathcal{M}=2$ & SU(n)*U(1) gauge field theory & & &
\\ \hline $\mathcal{N}=4$ and $\mathcal{M}=3$ with $n=3$ & Standard model& &
&\\ \hline
\end{tabular}
\caption{The correspondence between different quantum field theories in modern physics and the different composite knots in knot physics}
\end{table*}
\end{widetext}

We then discuss the Kelvin wave and knot dynamics on complex entangled
vortex-membranes -- composite knots. By projecting entangled vortex-membranes
into several coupled zero-lattices, the information of the system becomes the
coupled zero-lattices with internal degrees of freedom. After considering
\emph{the topological interplay between knots and the internal degrees of
freedom} (internal twistings, additional internal zeroes) -- twist-writhe
locking condition\cite{kou3}, different kinds of gauge interactions emerge. In
particular, it is the 3D quantum gauge field theories that characterize the
knot dynamics of the composite knots. The knot physics may give a complete
interpretation on quantum chromodynamics (QCD) and quantum electrodynamics
(QED). The collective motions of composite knots are described by quantum
fluctuations of zero-lattices: different fermionic elementary particles are
zeroes that are topological defects of different zero-lattices with
internal-twistings; \textrm{U(1)} gauge field (Maxwell field) are phase
fluctuations of internal twistings of internal T-type zero-lattice;
\textrm{SU(N)} Yang-Mills field are fluctuations of additional internal
zeroes, ... Two composite knots interact by exchanging fluctuations of the
internal-twistings. In table.1, we show the correspondence between different
quantum field theories in modern physics and the different composite knots
in\ knot physics.

A 3D $3$-level composite knot with ($\mathcal{N}=4$, $\mathcal{M}=3$) is an
object of 3D four entangled vortex-membranes $\mathrm{A}_{1},$ $\mathrm{A}%
_{2}$ and $\mathrm{B}_{1},$ $\mathrm{B}_{2}$ in 5D space. Here, $\mathcal{N}$
denotes the number vortex-membranes and $\mathcal{M}$ denotes the number of
levels. To generate a $3$-level composite knot with ($\mathcal{N}=4$,
$\mathcal{M}=3$), we firstly tangle two symmetric vortex-membranes
$\mathrm{A}_{1}$ and $\mathrm{A}_{2}$ and get \textrm{A}-knot. Next, we tangle
two symmetric vortex-membranes $\mathrm{B}_{1}$ and $\mathrm{B}_{2}$ and get
\textrm{B}-knot. Then, we tangle \textrm{A}-knot and \textrm{B}-knot into a
2-level composite knot with ($\mathcal{N}=4$, $\mathcal{M}=2$). Finally, we
symmetrically wind the 2-level knot with ($\mathcal{N}=4$, $\mathcal{M}=2$)
along different spatial directions and get a 3D $3$-level composite knot with
($\mathcal{N}=4$, $\mathcal{M}=3$). Fig.3(b) is an illustration of a composite
knot with ($\mathcal{N}=4$, $\mathcal{M}=3$). For the $3$-level composite
knot, there exist level-1 T/W-type zero-lattices, level-2 T/W-type
zero-lattices and level-3 W-type zero-lattices.

\begin{figure}[ptb]
\includegraphics[clip,width=0.52\textwidth]{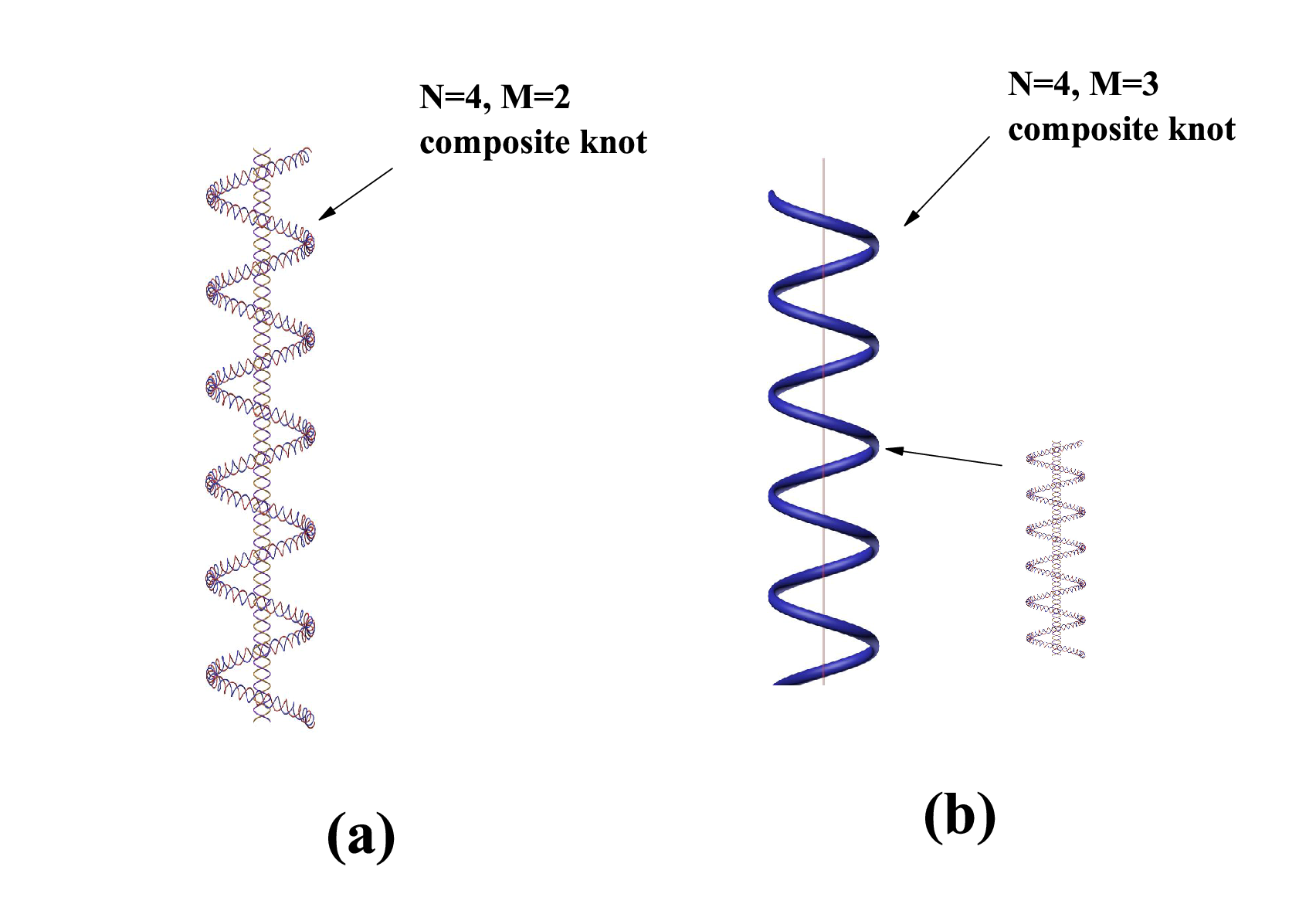}\caption{(a) The
illustration of 1D $2$-level composite knot with ($\mathcal{N}=4$,
$\mathcal{M}=2$); (b) The illustration of 1D $3$-level composite knot with
($\mathcal{N}=4$, $\mathcal{M}=3$). To generate a 1D 3-level composite knot
with ($\mathcal{N}=4$, $\mathcal{M}=3$), we symmetrically wind a 1D 2-level
knot with ($\mathcal{N}=4$, $\mathcal{M}=2$). }%
\end{figure}

By trapping different types of level-1 internal zeroes, there exist different
types of fermionic elementary particles. We use the following number series to
label different types of zeroes, $[n_{\mathrm{L}_{2}},$ $n_{\mathrm{L}_{1}}]$
where $n_{\mathrm{L}_{2}}$ is the half linking number of level-2 entangled
knot-crystals that is equal to the sum of the number of level-2 T-type zeroes
$n_{\mathrm{L}_{2}\mathrm{T}}$ and the number of level-3 W-type zeroes
$n_{\mathrm{L}_{3}\mathrm{W}}$ as $n_{\mathrm{L}_{2}}=n_{\mathrm{L}%
_{2}\mathrm{T}}+n_{\mathrm{L}_{3}\mathrm{W}},n_{\mathrm{L}_{1}}$ is the half
linking number of level-1 entangled vortex-membranes that is equal to the sum
of the number of level-1 T-type zeroes $n_{\mathrm{L}_{1}\mathrm{T}}$ and the
number of level-2 W-type zeroes $n_{\mathrm{L}_{2}\mathrm{W}},$ $n_{\mathrm{L}%
_{1}}=n_{\mathrm{L}_{1}\mathrm{T}}+n_{\mathrm{L}_{2}\mathrm{W}}.$ For the case
of $n=3$, the correspondence between four types of elementary particles is
given by
\begin{align*}
\text{An electron}  &  \Longleftrightarrow \text{A composite zero with
}[1,3]\text{,}\\
\text{A neutrino}  &  \Longleftrightarrow \text{A composite zero with
}[1,0]\text{,}\\
\text{An d-qurak}  &  \Longleftrightarrow \text{A composite zero with
}[1,1]\text{,}\\
\text{An u-qurak}  &  \Longleftrightarrow \text{A composite zero with
}[1,2]\text{.}%
\end{align*}

From point view of fluctuations of level-1 internal twistings, without
considering the fluctuations of level-3 W-type windings, the effective model
is an $\mathrm{SU_{\mathrm{strong}}(3)}\otimes$\textrm{U}$_{\mathrm{EM}}%
$\textrm{(1)} gauge theory. The \textrm{U(1)} gauge symmetry comes from
indistinguishable phase of 1-level internal twistings inside a composite zero.
Because the exact initial phase of a 1-level internal zero inside a composite
zero is not a physical observable value under projection, different choices of
initial phase angle of internal twistings lead to same physics result. The
\textrm{U}$_{\mathrm{EM}}$\textrm{(1)} gauge field characterizes the
interaction from the phase fluctuations of internal twistings. The
\textrm{SU}$\mathrm{_{\mathrm{strong}}(}$\textrm{3)} gauge symmetry comes from
indistinguishable states of $3$ 1-level internal-zeroes inside the composite
knot. The $\mathrm{SU_{\mathrm{strong}}(3)}$ gauge field characterizes the
interaction from the fluctuations of additional 1-level internal zeroes. If we
set the electric charge for an electron to be $e_{0}$, the electric charge for
an internal zero is $\frac{e_{0}}{n}.$ For the case of $n=3$, the charge of an
electron with $[1,$ $3]$ is $e_{0};$ the charge of a quark with $[1,$ $2]$ is
$\frac{2e_{0}}{3},$ the charge of a quark with $[1,$ $1]$ is $\frac{e_{0}}{3}%
$, the charge of a neutrino with $[1,$ $0]$ is $0.$ Because electron and
neutrino have no additional internal zero, the fluctuations of \textrm{SU}%
$\mathrm{_{\mathrm{strong}}}$\textrm{(3)} gauge field will never affect
electron and neutrino.

After considering the winding fluctuations of level-3 composite knot, the
electro-weak $\mathrm{SU}_{\mathrm{weak}}\mathrm{(2)\otimes U}_{\mathrm{Y}%
}\mathrm{(1)}$ gauge fields emerge. Electro-weak $\mathrm{SU}_{\mathrm{weak}%
}\mathrm{(2)\otimes U}_{\mathrm{Y}}\mathrm{(1)}$ gauge symmetry comes from the
redundancies of 2-level composite zeroes and those of 1-level internal zeroes
in unit cell of level-3 W-type zero-lattices. The fluctuations of
$\mathrm{SU}_{\mathrm{weak}}\mathrm{(2)}$ gauge theory come from the winding
fluctuations of level-3 composite knot and the fluctuations of $\mathrm{U}%
_{\mathrm{Y}}\mathrm{(1)}$ gauge theory come from the fluctuations of level-1
internal twistings (the internal zeroes) on a unit cell of level-3 W-type
zero-lattice. As a result, because the $\mathrm{SU}_{\mathrm{weak}%
}\mathrm{(2)}$ gauge field characterizes the interaction by exchanging the
winding fluctuations of level-3 composite knot, the $\mathrm{SU}%
_{\mathrm{weak}}\mathrm{(2)}$ gauge fields only couple to the
left-handed\emph{ }components of the lepton fields.

The effect of leapfrogging motion is to change a left-hand zero to a
right-hand zero. As a result, the fluctuating leapfrogging angular velocity of
a \emph{Standard knot} couples electro-weak $\mathrm{SU}_{\mathrm{weak}%
}\mathrm{(2)\otimes U}_{\mathrm{Y}}\mathrm{(1)}$ gauge fields and plays the
role of Higgs field in Standard model. The finite angular velocity of
leapfrogging motion plays the role of Higgs condensation and the Higgs
mechanism of 3-level composite knot-crystal with ($\mathcal{N}=4$,
$\mathcal{M}=3$) breaks the original gauge symmetry according to
$\mathrm{SU}_{\mathrm{weak}}\mathrm{(2)\otimes U}_{\mathrm{Y}}\mathrm{(1)}%
\rightarrow \mathrm{U}_{\mathrm{EM}}\mathrm{(1)}$.\ 

Finally, we derive the unified theory of Standard knot by considering all
points of view. The low energy effective theory is just the Standard model --
an $\mathrm{SU_{\mathrm{strong}}(3)}\otimes \mathrm{SU}_{\mathrm{weak}%
}\mathrm{(2)}\otimes$\textrm{U}$_{\mathrm{Y}}$\textrm{(1)} gauge theory with
Higgs mechanism due to spontaneous symmetry breaking. This is why I call this
composite knot with ($\mathcal{N}=4$, $\mathcal{M}=3$) to be the Standard knot.

\section{Emergent gravity}

\begin{figure}[ptb]
\includegraphics[clip,width=0.5\textwidth]{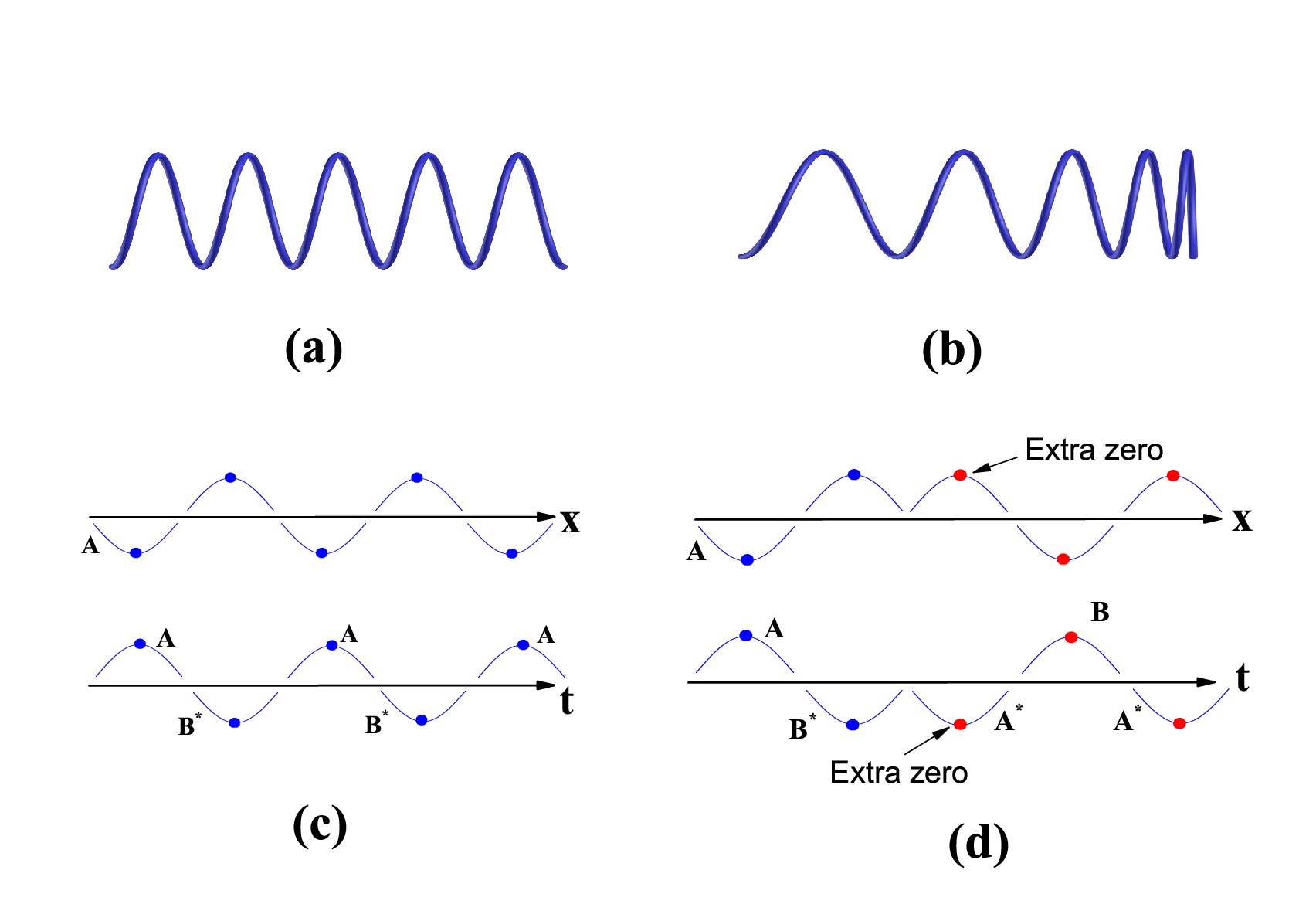}\caption{(a) The uniform
1D knot that corresponds to flat space-time for zeroes; (b) The deformed 1D
knot that corresponds to curved space-time for zeroes; (c) The illustration of
entanglement pattern of a uniform 1D knot with leapfrogging motion; (d) The
illustration of entanglement pattern of a 1D leapfrogging knot with an
additional zero that is a topological defect. }%
\end{figure}

In knot physics, there exists \emph{topological interplay effect between
zeroes and the knots}: on the one hand, the deformation of the knot leads to
the changes of particle motions that can be denoted by curved space-time; on
the other hand, the zeroes trapping topological defects deform the knot that
indicates matter may curve space-time. The Einstein-Hilbert action becomes a
topological mutual BF term that exactly reproduces the low energy physics of
the general relativity. We point out that to characterize the entanglement
evolution, the corresponding Biot-Savart mechanics for a zero on smoothly
deformed knot is mapped to quantum mechanics for particles on a curved
space-time. Fig.4(b) is an illustration of deformed 1D knot that corresponds
to curved space-time for matter.

For uniform entangled vortex-membranes, the knot is uniform and the scales are
(half) winding numbers along spatial/tempo directions. The deformation of the
knot leads to deformation of entanglement pattern. In particular, there exists
the diffeomorphism invariance that comes from information invariance of zeroes
-- when the lattice-distance of zero-lattice changes, the size of the zeroes
correspondingly changes.

There are two different descriptions to characterize the deformation of the
knot. On the one hand, to characterize the changes of the positions of zeroes,
we must consider a curved space-time by using geometric description; On the
other hand, we can introduce an auxiliary gauge field and use a gauge
description to characterize the deformation of the knot. There exists an
intrinsic relationship between gauge description for the deformation between
two vortex-membranes and geometric description for global coordinate
transformation of the same deformed knot.

From point view of geometry, a zero becomes topological defect of the knot --
an extra zero is not only anti-phase changing along arbitrary spatial
direction but also becomes anti-phase changing along tempo direction. Fig.4(d)
is an illustration of 1D deformed knot with an additional knot. From the point
view of gauge description, a knot traps a "magnetic charge" of the auxiliary
gauge field. In the path-integral formulation, to enforce such topological
constraint, we may add a topological mutual BF term in the action that is just
the Einstein-Hilbert action. The variation of the action via the metric gives
the Einstein's equations.

\section{Conclusion}

In summary, knot physics is developed to understand our
universe\cite{kou1,kou2,kou3}. The key point of knot physics is the ultimate
unification of everything (including matter, motion, interaction and
space-time, ...) into the entangled vortex-membranes that is a knot, i.e.,
\begin{align*}
&  \text{Everything in the universe}\\
&  \iff \text{Entangled vortex-membranes = A knot .}%
\end{align*}
In knot physics, our universe is a projection of Standard knot, a periodic
entanglement-pattern between four 3D vortex-membranes. All fundamental
interactions are unified into this simple framework, of which the low energy
effective theory not only reproduces the Standard model -- an
$\mathrm{SU_{\mathrm{strong}}(3)}\otimes \mathrm{SU}_{\mathrm{weak}%
}\mathrm{(2)}\otimes$\textrm{U}$_{\mathrm{Y}}$\textrm{(1)} gauge theory but
also leads to the physics of general relativity. In addition, the new theory
yields a deep understanding of the all kinds of elementary particles within
different structures of knots.


\begin{thebibliography}{99}                                                                                               %


\bibitem {stand}C. Quigg, \textit{Gauge Theories of the Strong, Weak, and
Electromagnetc Interactions,} Addison--Wesley Pub. Co., Menlo Park, (1983).

\bibitem {ss}M. Kaku, \textit{Introduction to Superstring and M-Theory 2nd
edition}. New York, Springer-Verlag (1999).

\bibitem {wen}{X.-G. Wen}, \textit{Quantum Field Theory of Many-Body Systems},
(Oxford Univ. Press, Oxford, 2004).

\bibitem {loop}C. Rovelli, arXiv:1102.3660.

\bibitem {e8}A. G. Lisi, arXiv:0711.0770 [hep-th] (2007).

\bibitem {Strand}C. Schiller, \textit{A fascinating speculation: The strand
model}.

\bibitem {twistor}R. Penrose, J. Math. Phys. 8, 345 (1967).

\bibitem {vo}Grigory E. Volovik, \textit{The universe in a helium droplet},
(Clarendon Press Oxford University Press New York, 2003).

\bibitem {kou1}S. P. Kou, Int. J Mod. Phys. B, V31, 1750241(2017).

\bibitem {kou2}S. P. Kou, Int. J Mod. Phys. B, V32, 1850090 (2018).

\bibitem {kou3}S. P. Kou, arxiv: 1706.06879, to be published as a chapter in
book \textquotedblleft \textit{Superfluids and Superconductors}%
\textquotedblright.

\bibitem {leap}Boris Khesin, Mosc. Math. J. 12 413 (2012).

\bibitem {1}N. Hietala, R. H\"{a}nninen, H. Salman, C. F. Barenghi, Phys. Rev.
Fluids 1, 084501 (2016).
\end{thebibliography}
\end{document}